\documentclass[aps,twocolumn,floatfix,superscriptaddress]{revtex4}
\usepackage[dvips]{graphicx,rotating,epsfig}
\usepackage{indentfirst}
\usepackage{amsmath,amssymb}
\usepackage{times}
\begin{document}

\title{An Interaction Potential for Atomic Simulations of Conventional High Explosives}

\author{Andrew J. Heim}

\affiliation{Department of Applied Science,
University of California, Davis, California 95616}

\affiliation{Theoretical Division,
Los Alamos National Laboratory, Los Alamos, New Mexico, 87545}

\author{Niels Gr{\o}nbech-Jensen}

\affiliation{Department of Applied Science,
University of California, Davis, California 95616}

\author{Edward M. Kober}

\author{Jerome J. Erpenbeck}

\author{Timothy C. Germann}

\affiliation{Theoretical Division,
Los Alamos National Laboratory, Los Alamos, New Mexico, 87545}

\date{\today}

\begin{abstract}

In an effort to develop a chemically reactive interaction
potential suitable for application to the study of conventional,
organic explosives, we have modified the diatomic AB potential
of Brenner \emph{et al}.~such that it exhibits improved detonation
characteristics. In particular, equilibrium molecular dynamics (MD)
calculations of the modified potential demonstrate that the detonation 
products have an
essentially diatomic, rather than polymeric, composition and that 
the detonation Hugoniot has the classic, concave-upward form.
Non-equilibrium MD calculations reveal the separation of scales
between chemical and hydrodynamic effects essential to the Zel'dovitch,
von Neumann, and D\"{o}ring theory.

\end{abstract}

\maketitle

\section{Introduction}

For over 15 years, the Reactive Empirical Bond-Order (REBO) potential of 
Brenner \emph{et al.} \cite{Brenner}, or variations thereof, have been used 
for Molecular Dynamics (MD) simulations of detonation. It has been shown to 
follow the Chapman-Jouguet (CJ) theory and even that of Zel'dovich, von 
Neumann, and D\"{o}ring (ZND)~\cite{Fickett}. 
Being an MD model, it has at least two major shortcomings, namely its spatial 
and temporal scales. As computers become more powerful and codes and 
algorithms become more advanced, some of the restrictions that 
limit these scales can be loosened, allowing for more realistic behavior 
to be modeled. 

While providing an atomic scale model of a reactive material, the empirical 
and classical model of Brenner \emph{et al}.~falls short of 
including all aspects of an accurate representation. Additionally, the spatial 
and temporal scales of real explosions make MD simulations a large 
computational task. 
But MD is still a useful tool for probing 
the characteristics of the detonation phenomenon, and the REBO potential 
is one of the best at balancing realism in the potential with accessibility to 
the large-scale; yet there is room for improvement. 

MD has certain conceptual advantages over hydrodynamics approaches that are 
parameterized to match the behavior 
of real high explosives. Even though the latter models can mimic real 
experiments on the proper 
spatial and temporal scales, they make assumptions about the reaction rate 
and multiphase equation of state (EOS) to do so. Some of these assumptions 
are parameter fittings, which may not elucidate any new physical 
intuition 
about detonation of the high explosive (HE) in question. In comparison, MD 
simulations depend on the parameterization of an interaction potential, 
which is arguably easier to connect directly to physical considerations 
than is a multiphase and multi-species EOS\@.

REBO has been used by many groups to model a variety of parameterizations and 
experimental configurations. However, a major criticism of this potential 
is that it has a thin reaction zone ($\sim 100$~\AA) relative to typical 
real high explosives ($\sim 1$~mm) \cite{Mader}. 
In previous work \cite{Heim,Swanson} 
several unconventional characteristics of the default parameterization of 
Brenner \emph{et al}.~\cite{Brenner} (called ModelI as in \cite{Swanson}) 
are made evident: 
\begin{itemize}
\item ModelI displays nearly instantaneous dissociation upon compression by an 
unsupported detonation. Its entire reaction zone is characterized by a 
dissociative state. 
\item Its reaction  zone is thin.  
\item It readily allows for clustering. 
\item Its CJ state is highly compressed.
\item It has clearly non-hyperbolic equilibrium Hugoniots ($\mathcal{H}$) 
in $P$--$v$ space. 
\end{itemize}

None of these characteristics are 
proven to be unrealistic. In fact for some primary high explosives,   
the plasma-like state seen in ModelI at CJ may be realistic \cite{Heflinger}. 
Clustering, particularly of carbon, is a real phenomenon in the thermal 
decomposition of some HEs \cite{Tao,Greiner}.  
However, successful models of conventional 
explosives have either assumed or suggested that more molecular states 
\cite{Fried, Howard} with hyperbolic $\mathcal{H}$s in $P$--$v$ space 
\cite{Mader, Ree, Shaw2} are typical and that compressions at CJ conditions 
should be $\approx 25\%$ (for example PETN \cite{Ree}) 
rather than $\approx 43\%$ as with ModelI \cite{Heim}. 
Given that ModelI can model a dissociated state, it would be useful to 
show that a more molecular state can also be modeled. The task of this work 
is to make modifications to the ModelI potential in order to accomplish this 
goal.   
Although, for the last feature in the above list, it is 
difficult to predict what changes would adjust it, an attempt to address the 
former features is made by making adjustments based on physical reasoning (see 
Sec.~\ref{sec:fix}). In Sec.~\ref{sec:compare} the 
effects that the changes to the model of Brenner \emph{et al.}~have on the 
thermodynamic properties are investigated by the methods outlined in 
Sec.~\ref{sec:meth}. 
Notable results of non-equilibrium MD simulations are studied in the companion 
paper.

\section{Changing REBO: physical and aesthetic reasoning}
\label{sec:fix}

In this section we motivate changes to the ModelI potential to form a new 
potential (called ModelIV after the naming scheme of White 
\emph{et al.}~\cite{Swanson}), the total bonding energy of which takes the form
\begin{equation}\label{eq:JR}
\begin{split}
E_b=\sum_{i}^N\sum_{j>i}^N\{&f_{c}(r_{ij})[(2-\overline{B}_{ij})
V_R(r_{ij})-\overline{B}_{ij}V_{A}(r_{ij})]+\\&V_{vdW}(r_{ij})\}. 
\end{split}
\end{equation}
The parameters for Eq.\ref{eq:JR} can be found in Table~\ref{tab:JR}\@.
\begin{table}
\caption{The components and parameters used in Eq.\ \ref{eq:JR}.}
\label{tab:JR}
\begin{eqnarray*}
V_{R}(r)&=&\frac{\mathcal{D}_{e}}{S-1}\exp\left[-\beta\sqrt{2S}(r-r_{e})\right]\\
V_{A}(r)&=&\frac{S\mathcal{D}_{e}}{S-1}\exp\left[-\beta\sqrt{\frac{2}{S}}(r-r_{e})\right]\\
B_{ij}&=&\left\{1+G\sum_{k\neq i,j}f_{c}(r_{ik})\exp[m(r_{ij}-r_{ik})]\right\}^{-n}\\
y(r)&=&\frac{r-\gamma_1}{\gamma_2-\gamma_1}\\
f_{c}(r)&=& \left \{ 
  \begin{array}{lrcl}
    1 &  &y(r)&<0\\ 
    (1-y(r))^3 (1+3y(r)+6y^2 (r)) & 0 \le &y(r)& < 1\\
    0 & 1 \le &y(r)&
\end{array} 
\right.  \\
V_{vdW}(r)&=&\left\{ 
  \begin{array}{lrcl}
    \epsilon c & &r& < \alpha_1\\
    \epsilon \left(c + \displaystyle\sum_{i=3}^5 P_i (r-\alpha _1)^i \right) & \alpha_1 \le &r& < \alpha_2\\
    \epsilon \left[\left(\frac{r^*}{r}\right)^{12}-2\left(\frac{r^*}{r}\right)^{6}\right] & \alpha_2 \le &r& < \alpha_3\\
    \epsilon (C_0+C_1r+C_2r^2+C_3r^3) & \alpha_3 \le &r& < \alpha_4\\
    0 & \alpha_4 \le &r&
  \end{array}
\right. \\
\end{eqnarray*}
\begin{tabular}{p{\columnwidth}}
\hline
\raggedright\scriptsize
$\textrm{Mass}_\textrm{A}=14.008\textrm{~amu};~\linebreak[0]
\textrm{Mass}_\textrm{B}=12.010\textrm{~amu};~\linebreak[0]
\epsilon=5.0\times10^{-3}\textrm{~eV};~\linebreak[0]
c=200;~\linebreak[0]
D_{e}^{\textrm{AA}}=D_{e}^{\textrm{BB}}=5.0\textrm{~eV};~\linebreak[0] 
D_{e}^{\textrm{AB}}=1.0\textrm{~eV};~\linebreak[0] 
\mathcal{D}_e = D_e + c \epsilon;~\linebreak[0]
S=1.8;~\linebreak[0]
\beta=2.7\textrm{~\AA}^{-1};~\linebreak[0] 
r_{e}=1.2\textrm{~\AA};~\linebreak[0] 
G=5.0;~\linebreak[0] 
m=2.25\textrm{~\AA}^{-1};~\linebreak[0] 
n=0.5;~\linebreak[0] 
\overline{B}_{ij}=\frac{1}{2}(B_{ij}+B_{ji});~\linebreak[0] 
\gamma_1=1.3~r_e;~\linebreak[0] 
\gamma_2=1.7~r_e;~\linebreak[0] 
r^*=2^{\frac{1}{6}}\times2.988\textrm{~\AA};~\linebreak[0] 
\alpha_2=2^{-\frac{1}{6}}r^*;~\linebreak[0] 
\alpha_1=0.683~\alpha_2;~\linebreak[0] 
\alpha_3=\left(\frac{13}{7}\right)^{\frac{1}{6}}r^*;~\linebreak[0] 
\alpha_4=\frac{67}{48}\alpha_3;~\linebreak[4] 
P_3=-2290.707325617024;~\linebreak[0] 
P_4=3603.929410034915;~\linebreak[0] 
P_5=-1513.930039501751;~\linebreak[4] 
C_0=2.575275778983429;~\linebreak[0] 
C_1=-4.316677142326428;~\linebreak[0] 
C_2=1.376573417835169;~\linebreak[0] 
C_3=-0.12340088128894569;~\linebreak[0]$
\end{tabular}
\end{table}
Before mentioning the physical basis for the modifications made to 
ModelI, let us list the features that ModelI already takes into account. 
In tests using ModelI, the parameterization is set to a 
valence of one, which 
should prefer dimers because one atom's sharing of electrons with one other 
fills both atoms' valence shells. The bonding energy takes the form  
\begin{equation}\label{eq:BREW}
E_b=\sum_{i}^N\sum_{j>i}^N\{f_{c}(r_{ij})[
V_R(r_{ij})-\overline{B}_{ij}V_{A}(r_{ij})]+V_{vdW}(r_{ij})\}.
\end{equation}
See Table I in the errata of Brenner \emph{et al}.~\cite{Brenner} for the 
functions and components of Eq.~\ref{eq:BREW}. 
In ModelI this bond-order 
behavior is implemented by reducing the strength of the attractive term of a 
Morse potential used to model such covalent bonding. 
This is accomplished with a bond-order coefficient ($\overline{B}_{ij}$) 
that varies from one to zero with increasing proximity and number of 
neighbors. 

Rice \emph{et al}.~found that ModelI allowed for trimer formation and 
subsequently strengthened this bond-order effect by 
increasing the repulsive term as well in order to better model bond saturation 
so that a particle became less likely to bond to more than one 
other \cite{Rice1,Rice2}\@. 
The contribution of Rice \emph{et al}.~is a simplified adjustment 
for what can be a complicated interaction. For instance, the 
$\overline{B}_{ij}$ coefficient does not take spin or degeneracy into account. 
However, this type of correction is similar to the "bond saturation" term that 
is used in the more sophisticated and calibrated ReaxFF potential of 
van~Duin \emph{et al}.~\cite{vanDuin}. 

In combination with this bond-order functionality, the variation of bond well 
depths allows for reactive chemistry. 
If $i$ and $k$ are different types of atoms, their respective well depths with 
$j$ will be different. If $j$'s is deeper with $k$ than with $i$, $k$ will 
require less kinetic energy to displace $i$ than $i$ would have, had their 
rolls been reversed. 
That difference in energy will be converted into the kinetic energy of the 
system, an exothermic reaction. This is the type of reaction modeled by ModelI, 
$2\textrm{AB} \rightarrow \textrm{A}_2 + \textrm{B}_2 + 2Q$, where 
$Q \equiv D_e^\textrm{AA/BB} - D_e^\textrm{AB}$ is the 
exothermicity of the reaction. The amount of energy 
needed to dissociate AB is $D_e^\textrm{AB}$\@. To dissociate AA or BB 
requires $D_e^\textrm{AA/BB}$\@. These numbers ignore the small 
contribution of the $V_{vdW}$ term in Eq.~\ref{eq:BREW}. 

\begin{figure}
\includegraphics*[width=\linewidth]{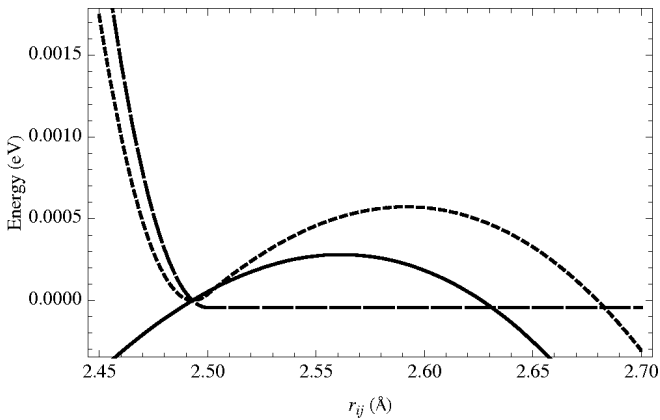}
\caption[Total bond energy in a 1D, three-particle, A-B-A, ModelI 
interaction in which the distance between the B and one of the A's is fixed at 
the bond's equilibrium distance vs.~the distance between B and the remaining A 
(short dash)\@.]
{Total bond energy in a 1D, three-particle, A-B-A, ModelI 
interaction in which the distance between the B and one of the A's is fixed at 
the bond's equilibrium distance vs.~the distance between B and the remaining A 
(short dash)\@. $V_{vdW}$ for any two atoms (solid)\@. 
Total bonding energy less the $V_{vdW}$ 
contributions (long dash)\@. All curves are subtracted by their corresponding 
values at the position of the local minimum of the first curve for comparison 
purposes. This shows that the section of positive slope in the REBO 
$V_{vdW}$ inner spline can trap particles. Without it, there is no trap.
\label{fig:brenner_trap}}
\end{figure}

In ModelI the van der Waals (vdW) interaction is represented by a 
Lennard--Jones (LJ) form. 
It is connected with a terminating third-order polynomial with negative 
curvature in the short range domain so that the overly repulsive 
twelfth-order term 
does not compete with the Morse potential, which handles the 
covalent bonding and Pauli Exclusion repulsion. The LJ is parameterized 
to allow for solid lattice formation at low temperature. 
It turns out that 
the third-order inner spline introduces an artifact in ModelI 
since it has a section of positive slope, which allows it to trap particles 
(see Fig.~\ref{fig:brenner_trap}) because of the resulting attractive force. 
It is only first-derivative 
continuous at the spline point and is cropped at the cutoff point. 

ModelI simulates well the repulsion between individual atoms separated by less 
than their equilibrium distance---which depends on the number and proximity of 
neighbors to the pair. It does not, however, represent the electrostatic 
repulsion between dimers that occurs as their charge clouds overlap on 
approaching each other while still being too far away to
rearrange bonds, $\sim3$~\AA. As a result there is no significant interaction 
between molecules at the range and magnitude that
gives rise to high pressure dense molecular HE product fluid mixtures 
\cite{Shaw}\@. 
There are some good models for this, for example, 
ZBL~\cite{ZBL} 
and ReaxFF~\cite{vanDuin}\@. 
In an attempt to improve upon ModelI and its 
results, we design a new model (ModelIV), which incorporates, along with the 
aforementioned contribution of Rice~\emph{et al.}, the following 
modifications: 

The inner spline is replaced with a repulsive core, 
which smoothly connects the 
negative (the zero of the potential energy is the dissociated state) 
LJ curve to a constant plateau in the region in which the Morse 
potential dominates. This inner core has no sections of positive 
slope---that is, sections of attractive force---and therefore 
cannot trap atoms, and it crudely models electrostatic repulsion. 

The reason for connecting to a 
plateau is that it allows one to dictate the depth of the bonding potential. 
It also simplifies the definition of a covalent bond. Any two particles that 
are within the defined bond distance (see the $f_c$ term in 
Table~\ref{tab:JR})  
with a radial component of kinetic 
energy less than the height of the plateau are considered bonded. This 
definition also distinguishes bonded particles from particles that are merely 
confined by neighbors, and it can be determined at any instant. 
The height of the plateau, 1~eV, is chosen to be comparable to the 
thermal energy at the CJ state for ModelI~\cite{Heim}, an addition that 
should 
decrease the reaction cross section, desensitizing the HE to initiation and, 
perhaps, adjusting reaction time and, thus, the width of the reaction zone.
Perhaps, a more physical function would continue to increase monotonically to 
a finite value at $r=0$ and have $V_{vdW}$ be particle-type dependent. 

\begin{figure}
     \includegraphics*[width=\linewidth]{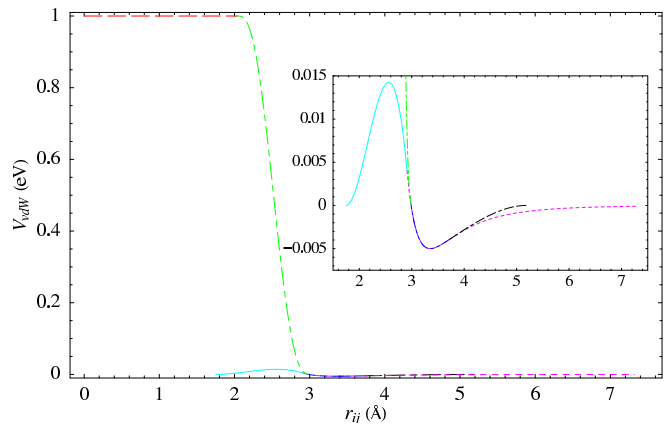}
\caption[van der Waals ($V_{vdW}$) potential vs.~the interatomic 
distance ($r_{ij}$)\@.]
{(Color online) van der Waals ($V_{vdW}$) potential vs.~the interatomic 
distance ($r_{ij}$)\@. Spline points are indicated by changes in 
color--dashing. 
ModelI $V_{vdW}$ is made of a third-order polynomial (cyan--solid) 
connected to a 
Lennard--Jones (LJ) form (mauve--short dash), which is cropped at a finite 
distance at which the slope and value of $V_{vdW}$ are nearly zero. 
ModelIV  $V_{vdW}$ starts as a constant (red--long dash) and 
is connected  with a fifth-order polynomial (green--long short dashes) 
so that the spline points are second-derivative 
smooth, to a LJ form (blue--short short long dashes) at zero, 
below which it coincides with the ModelI LJ until the inflection 
point, at which it is connected to a curve (black--long long short dashes) 
that is brought smoothly to zero within a finite distance, a Holian--Evans 
spline \cite{Evans}\@.
\label{fig:vdW_cmprsn}}
\end{figure}

\begin{figure}
      \includegraphics*[width=\linewidth]{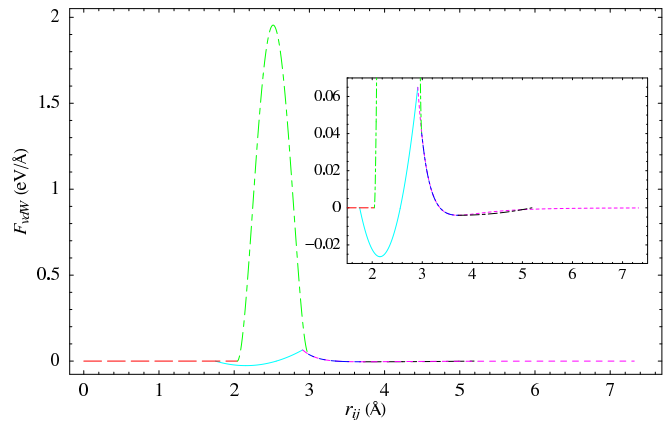}
\caption{(Color online) Forces that correspond to the potentials in 
Fig.~\ref{fig:vdW_cmprsn}.
\label{fig:vdW_f_cmprsn}}
\end{figure}

All of the spline points in ModelIV have, 
at least, continuous second derivatives, with the 
exception of the outermost cutoff point. To terminate the $V_{vdW}$ term in 
the long range, a Holian--Evans spline \cite{Evans}, which is second 
derivative continuous at its inner spline point but only first derivative 
continuous at its outer one, is used. A computational advantage 
for using the Holian--Evans 
spline is that the reach of the ModelIV potential is shorter 
($\approx 5.19$~\AA) than ModelI's ($\approx 7.32$~\AA)\@.  
Fig.~\ref{fig:vdW_cmprsn} shows a comparison of the two vdW potentials and 
Fig.~\ref{fig:vdW_f_cmprsn} shows the corresponding forces. One can see that 
the ModelIV $V_{vdW}$ term is still smooth in the force.
As with the spline points in the $V_{vdW}$ term, the cutoff function in the 
bond-order function and Morse potential is also replaced so that it is 
second-order continuous. 
Second derivative continuity helps energy conservation in 
the MD simulations (and allows longer time steps to be taken).  

\begin{figure}
\includegraphics*[width=\linewidth]{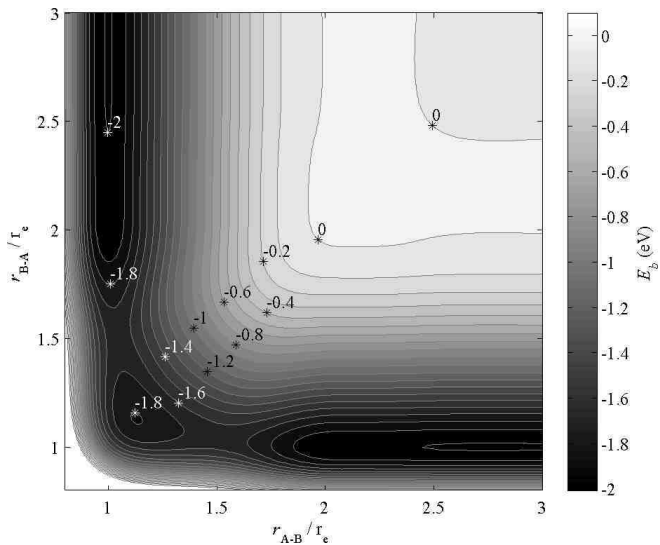}
\caption[Potential energy surface, ModelI]
{Potential energy surface for the linear, symmetric, metastable configuration 
of ModelI atoms. The contours are spaced every 0.1~eV\@. $r_{\textrm{A--B}}$ 
is the distance between and A and B atom. $r_{\textrm{B--A}}$ is the the 
distance between the same A atom and a different B atom. $r_e$ is the 
equilibrium distance for the metastable covalent interaction. 
\label{fig:REBO_A-B-A}}
\end{figure}

\begin{figure}
\includegraphics*[width=\linewidth]{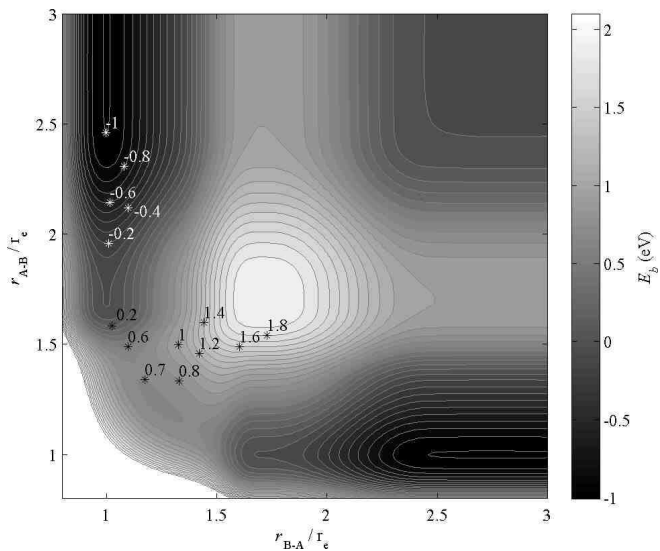}
\caption[Potential energy surface, ModelIV]
{Potential energy surface for the linear, symmetric, metastable configuration 
of ModelIV atoms. The contours are spaced every 0.1~eV\@.
\label{fig:JR_A-B-A}}
\end{figure}

Another difference between ModelIV and ModelI is the depth of 
the metastable covalent wells. With the addition of the repulsive core, 
ModelIV fails to detonate with a well as deep as  
the default value for ModelI. 
The metastable well is raised by 1~eV, such that $D_e^{\textrm{AB}}=1.0$~eV, 
$D_e^{\textrm{AA/BB}}=5.0$~eV, and therefore, $Q=4.0$~eV\@.
Comparing the potential energy surface for the linear, symmetric, 
metastable configuration for ModelI (see Fig.~\ref{fig:REBO_A-B-A}) to the one 
for ModelIV (see Fig.~\ref{fig:JR_A-B-A}), we notice that the depression near 
(1.2 $r_e$, 1.2 $r_e$) diminishes somewhat for ModelIV and that the barrier to 
be overcome for one particle to displace another via an end-on attack is 
significantly increased.
All of these changes have notable effects on the EOS and 
detonation properties of the model HE\@. The final difference is that 
the masses of the particles are changed so that they are different from 
each other. The changes to ModelI to form ModelIV may be summarized as follows:
\begin{itemize}
\item A bond-order coefficient is applied to the repulsive term in the Morse 
part of the interaction potential.
\item The inner spline of the $V_{vdW}$ term is replaced with a 
repulsive core, which comprises a plateau connected to an LJ  
form with a monotonically decreasing fifth-order polynomial.
\item The order of all of the splines is increased such that all but one 
spline point (the outermost cutoff) have smooth  second derivatives. 
\item The masses of the atom types are changed so that they are no longer 
equal. 
\item The depth of the binding energy for the metastable bonds is made 
shallower.
\end{itemize}

\section{Methods}
\label{sec:meth}

Investigating the two REBO potentials of which this paper is a 
study, ModelI and ModelIV, we use SPaSM~3.0 
\cite{Lomdahl} to carry out the MD simulations, of which there are two main 
types, microcanonical ensembles (NVE) and non-equilibrium MD (NEMD)\@. The NVE 
simulations are used for testing the thermochemical properties of a model and 
NEMD are used to investigate its detonation behavior. Both utilize the 
leapfrog Verlet method to advance the atoms' positions and momenta. 

We conduct two sub-types of 
NVE simulations. The first is meant to find the equilibrium Hugoniots and are 
similar to those described by others \cite{Heim,Rice1,Erpenbeck}. 
The new model requires a 
shorter time step $\delta t \approx 0.16$~fs than does that of Brenner 
\emph{et al.}~\cite{Brenner}, $\delta t \approx 0.25$~fs because of the 
relatively high curvature of the fifth-order spline in ModelIV's $V_{vdW}$ 
term. Its zero-pressure configuration 
is also slightly different. The length of the sides of the rectangular unit 
lattice cell are  
$l_x=6.19122$~\AA~and $l_z = 4.20538$~\AA\@. The cell contains two dimers 
in a herringbone configuration. 
The angles that the dimers make with the horizontal are  
$\pm 27.7109^\circ$. The atoms are placed 0.59976~\AA~from the centers of 
their respective dimers, which are positioned at (1/4~$l_z$, 1/4~$l_x$) and 
(3/4 $l_z$, 3/4~$l_x$) from the lower left corner of the cell.  
These values are determined by isothermal--isobaric 
Monte Carlo simulations. 

The second type of NVE simulation, the cookoff, 
is used to 
determine the reaction rate. The initial conditions are the same as with the 
first type of NVE except that the constituents are 100$\times$100 cells$^2$ 
of AB not 25$\times$25 of AA and BB\@. The time step 
is also drastically reduced so that good statistics can be found for 
measurements taken during rapid reactions. The time step varies among 
cookoff simulations, 
$6 \times 10^{-4}$~fs~$< \delta t < 0.02$~fs. 

One of the purposes of creating a new potential is to expand the reaction zone 
by lowering the reaction cross section. This requires larger NEMD simulations 
to be performed. As Rice \emph{et al.}~have done~\cite{Rice1} for computational 
efficiency, we 
try to reduce the amount of pre-shocked material modeled by tacking initial 
state material onto the end of the sample as the shock front approaches. To 
reduce surface effects, we introduce at the surface a one-cell-thick layer of 
two new particles, C and D, that have all 
of the properties of A and B, respectively, except that they start frozen and 
their velocities are not updated. When the shock front nears the frozen layer, 
the frozen layer is converted to A and B 
and is thermalized along with the new material. Another frozen layer is tacked 
onto the end. The size of an NEMD simulation is discretely dependent on time. 
The largest simulation contained over three million atoms. 

Another effect of decreasing the reaction cross section is that the incubation 
time, before which detonation occurs, extends. The sensitivity to impact is 
thus lessened. To overcome this, we initially overdrive the simulation with a 
piston that moves into the material at a velocity $u_{pstn} \approx 4.9$~km/s. 
After a detonation front has seemingly been established, it is smoothly backed 
off by following a sine-shaped deceleration over 200 time steps to a desired 
velocity. The time step is the same as with the first type of NVE simulation, 
$\delta t \approx 0.16$~fs. Most of the results of the NEMD simulations are 
reported in the companion paper. 

\section{Comparison of Thermodynamic Properties}
\label{sec:compare}

We start the comparison of the properties of ModelI and ModelIV by analyzing 
how the ModelIV material behaves when compressed by the passage of an 
under-driven detonation front. 
\begin{figure}
\includegraphics*[width=\linewidth]{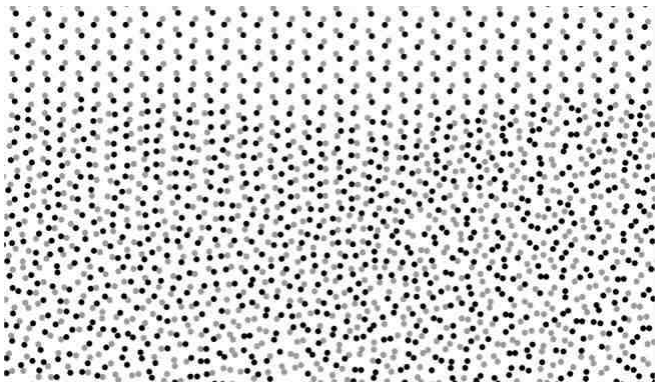}
\caption[Snapshot of a section of a shock front for an 
under-supported detonation using the ModelIV potential.]
{Snapshot of a section of a shock front for an 
  under-supported detonation using the ModelIV potential. Shock is 
propagating upward. Particles are shaded by atom type.}
\label{fig:snpsht_JR_frnt}
\end{figure}
From the snapshot of an NEMD simulation using ModelIV (see 
Fig.~\ref{fig:snpsht_JR_frnt}), one can see on the left half that many 
unreacted dimers are lining up horizontally. This is representative of 
uniaxial compression. Farther back it has melted, and farther back still 
reacted products are evident. 
On the right half, the dimers do not seem to line up 
before reacting. It is made evident in the companion article 
that this is an effect of detonation 
instability and the propagation of transverse waves. When comparing this to a 
snapshot of detonation in ModelI (see Fig.~1 in the paper of Heim 
\emph{et al}.~\cite{Heim}), 
one notices that, even at the right side of the sample, ModelIV seems to 
hold its molecular identity better than ModelI during compression by a 
detonation front that is not overdriven. It seemingly displays a greater 
resilience to dissociation. Both the right half of the ModelIV snapshot 
and the ModelI snapshot lack a significant induction zone. 

\begin{figure}
\includegraphics*[width=\linewidth]{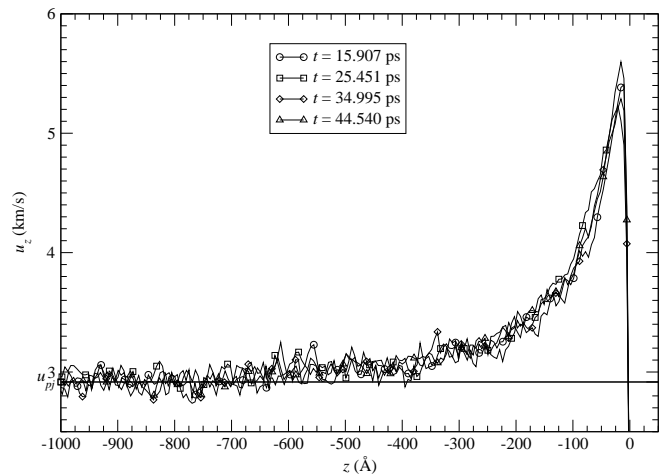}
\caption[Overlap of profiles of the $z$-component of the particle velocity 
for a critically supported NEMD simulation using ModelIV  
parameterized 
such that $D_e^{\textrm{AB}}=1.0$~eV and $Q=4.0$~eV\@.]
{Overlap of profiles of the $z$-component of the particle velocity 
for a critically supported NEMD simulation using ModelIV  
parameterized 
such that $D_e^{\textrm{AB}}=1.0$~eV and $Q=4.0$~eV\@. A constant line is 
drawn at the CJ value of the particle velocity ($u_{pj}$), 
at which the driving piston is moving. Using a critically supported NEMD as 
opposed to an unsupported one drastically reduces the transient time from 
initiation to steady state \cite{Heim}.}
\label{fig:vz_ovrlp_crit_JR}
\end{figure}

To better compare 
the widths of the detonation fronts of the two models, profiles of the 
average $z$-component of particle velocity from a critically 
supported detonation of ModelIV are plotted. 
Figure \ref{fig:vz_ovrlp_crit_JR} indicates 
that the reaction zone is about 700~\AA~wide since that is the position at 
which the curves settle down to the constant $u_{pj}$. 
With the same method, the width of the reaction zone for ModelI was determined 
to be about 300~\AA~\cite{Heim}. 

Although ModelIV's reaction zone is still small compared to that of real 
explosives, it can be widened. 
To widen the 
reaction zone farther, one can create a reaction that requires more 
steps. One also can have the reaction increase the number of moles of 
material. The resulting expansion in volume would be another driver for the 
shock wave \cite{Fickett}. Yet another effect of this would be to introduce an 
entropic penalty to back-reaction and make the model more closely 
approximate the ZND assumption of irreversibility. 

To find the critical piston velocity in the preceding analysis, 
the CJ state for ModelIV is found 
by conducting NVE simulations and seeking the values of $E$ and $v$ that 
satisfy the Hugoniot jump conditions \cite{Heim,Rice1,Erpenbeck}\@. 
For ModelI $v_j/v_0 \approx 0.57$ and $u_{sj} \approx 9.7$~km/s \cite{Heim}.
From the results (see 
Table~\ref{tab:CJ_JR}), one should notice that for ModelIV the CJ state is less 
compressed than for ModelI and that 
$u_{sj}$ is faster. Compared to 
conventional explosives, which typically have specific volumes 
$v_j/v_0\approx 0.75$ and shock velocities $u_{sj}\approx 6$~km/s 
(for example PETN \cite{Ree}), ModelIV 
detonation fronts are over twice as fast and its CJ state is slightly less 
compressed. Note, however, that it has been shown for these REBO potentials 
that 3D simulations have lower, and therefor more realistic,  velocities than 
do 2D ones \cite{Swanson,Maillet}. 
\begin{table}
\caption{Determined thermodynamic values at CJ for ModelIV\@, where $v$ is the 
specific volume, $u_s$ is the shock velocity, $u_p$ is the particle velocity, 
$T$ is the temperature, $U$ is the potential energy, $E$ is the internal 
energy, $P$ is the pressure, and $\lambda$ is the degree of reaction. 
Subscript $j$ indicates the CJ value, Subscript $0$ indicates the value at the 
initial state. Averages are per particle.}
\label{tab:CJ_JR}
\begin{tabular}{|l|l|}\hline
  $v_j/v_0$				&0.789851(39) \\ \hline
  $u_{sj}$				&13.87868(50) \\ \hline
  $u_{pj}$				&2.91658(44)  \\ \hline
  $\langle k_{B}T_j\rangle$~(eV)	&0.9185(84)   \\ \hline
  $\langle U_j\rangle$~(eV)		&-0.8631(72)  \\ \hline
  $\langle E_j \rangle$~(eV)		&0.0554(12)   \\ \hline
  $P_j$~(eV/\AA$^{2}$)			&0.83846(16)  \\ \hline
  $\lambda_j$				&0.8548516(36)\\ \hline
\end{tabular}
\end{table}
\begin{figure}
\includegraphics*[width=\linewidth]{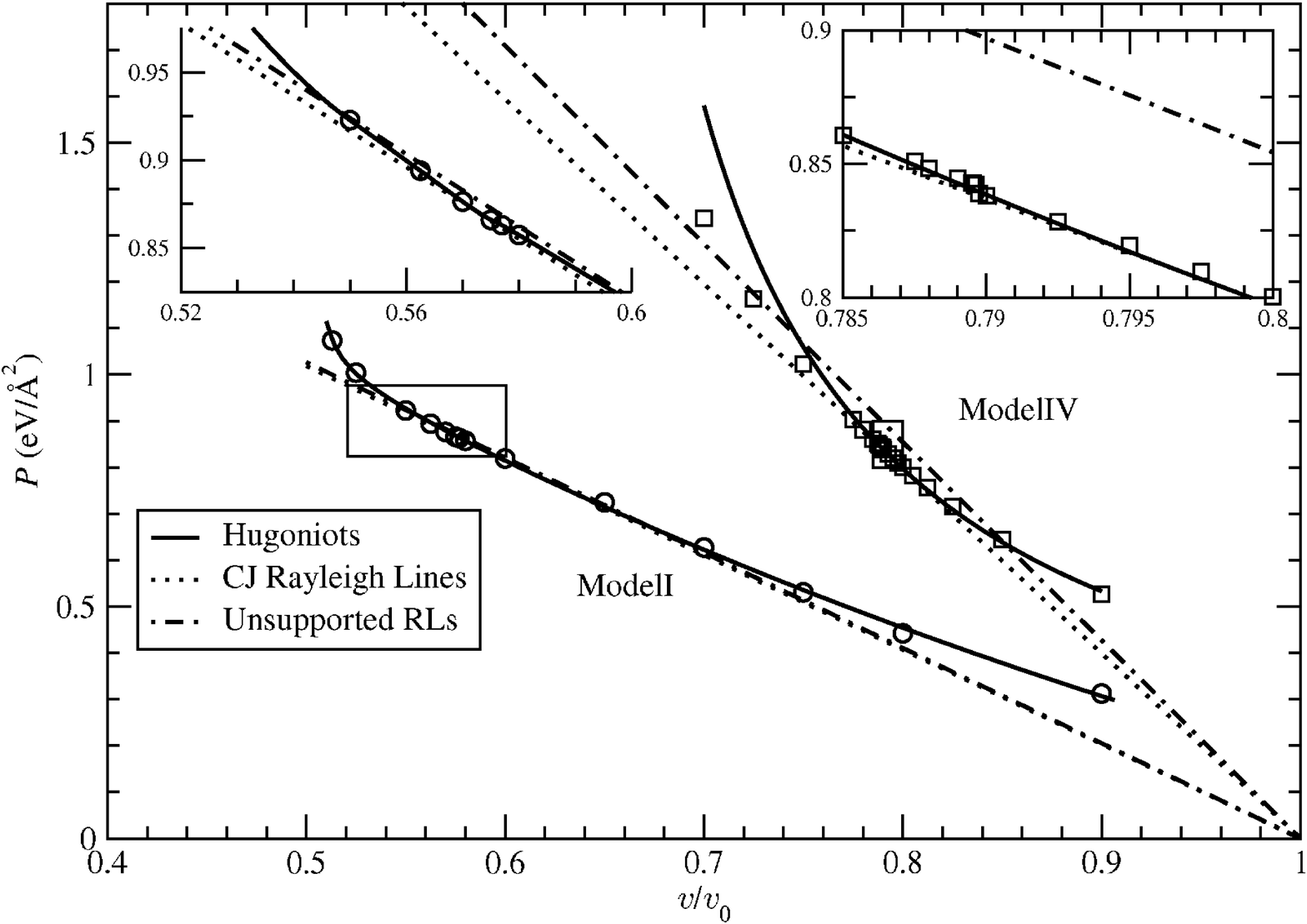}
\caption[Hugoniots for ModelI (circles) and ModelIV 
(squares)\@.]
{Hugoniots for ModelI (circles) and ModelIV 
(squares)\@. Solid lines are a guide to the eye for ModelI and a fit for 
ModelIV\@.  
Rayleigh lines are determined by the initial conditions and slopes. Dotted 
lines represent Rayleigh lines the slope of which are determined by the CJ 
value of the detonation velocity. Slopes of dash-dotted lines are 
determined by the average velocity of the shock waves in unsupported 
detonations. Boxes are magnifications. }
\label{fig:Pv_REBO_JR}
\end{figure}
Heim~\emph{et al.}~\cite{Heim} show  
how closely the conditions of the final 
state in a detonation of a ModelI material match those of the CJ 
conditions. 
For contrast in Fig.~\ref{fig:Pv_REBO_JR}, one can see that the Rayleigh line  
for the unsupported simulation is significantly steeper than for that of the 
CJ state. 

In ZND theory this would require that the final state for ModelIV is either at 
a strong point or a weak point. The strong point is unstable and will only 
be a solution to the conservation equations if the detonation is overdriven. 
In the strict ZND theory there is no path to the weak point. ZND is, 
however, 
based on assumptions not realized by ModelIV\@. ModelIV has a 
reversible 
reaction, the pathway of which includes an endothermic dissociation. 
Endothermic steps can be responsible for weak point final states 
\cite{Fickett}. Perhaps, the repulsive core introduced in ModelIV, by 
sheltering 
the dimers, prevents the reduction of the activation 
energy of the dissociative step and, hence, increases that step's reaction 
time, thus making it a stronger contributer to the overall reaction rate and 
causing the final state to be at a weak point. As is shown in the 
accompanying paper and can be inferred from Fig.~\ref{fig:snpsht_JR_frnt}, 
however, a 1D theory is not totally appropriate for this 
model.

\begin{figure}
\includegraphics*[width=\linewidth]{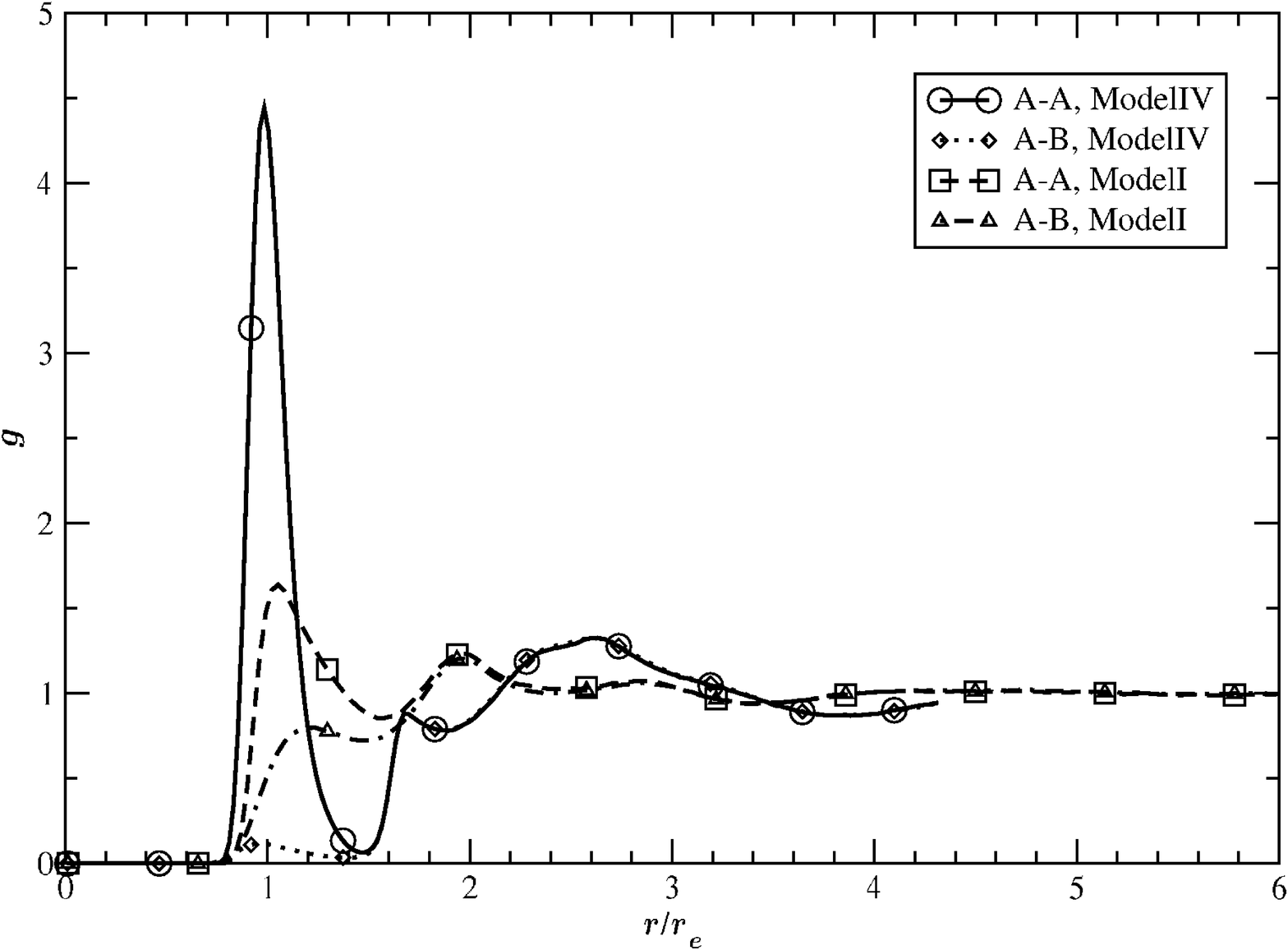}
\caption[Radial distribution functions for the CJ states of ModelIV, 
parameterized such that $D_e^{\textrm{AB}}=1.0$~eV and $Q=4.0$~eV, 
and of ModelI\@.]
{Radial distribution functions for the CJ states of ModelIV and of ModelI\@.}
\label{fig:RDF_BREWnJR}
\end{figure}

\begin{figure}
\includegraphics*[width=\linewidth]{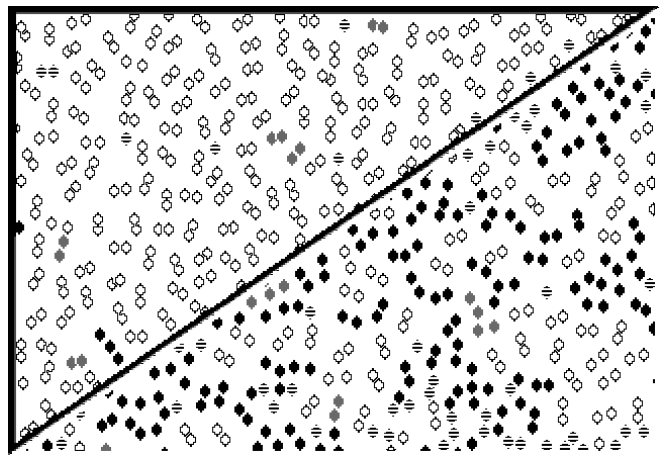}
\caption[Snapshots of the CJ state for ModelIV (upper) and ModelI (lower)]
{Snapshots of the CJ state for ModelIV (upper) and ModelI (lower)\@. 
Particles are marked by bond type. Gray atoms  
are in unreacted dimers; striped are unbonded, dissociated atoms; black are 
clustered atoms; and white are reacted dimers.}
\label{fig:snpshts_CJ}
\end{figure}

In order to confirm that the modifications to the REBO model in the ModelIV 
potential do, indeed, maintain a dimerized state, the radial 
distribution function (RDF) at the CJ state is plotted. 
In Fig.~\ref{fig:RDF_BREWnJR} 
the CJ state of ModelIV is compared to that of ModelI\@. From the RDF for 
ModelIV, one can 
see that, after the peak at $r=r_e$, the curve drops to nearly zero while the 
analogous region in the RDF of ModelI is closer to unity. This is indicative of 
a molecular state for ModelIV and a dissociative state for ModelI\@. From 
Fig.~\ref{fig:snpshts_CJ}, it is clear that the snapshot of the 
ModelIV CJ state supports the finding of the corresponding RDF\@. 
The snapshot for ModelI shows many more clusters and dissociated atoms. 

For these images, a bond is defined for each potential. Two particles, 
$i$ and $j$, are considered bonded if $E_{b,ij}+KE_{\parallel,ij} < E_c$ and 
$r_{ij} < r_c$, where $E_{b,ij}$ is as in Eq.~\ref{eq:JR} for ModelIV and 
Eq.~\ref{eq:BREW} 
for ModelI except that the outer sum is over only $i$ and $j$. 
$KE_{\parallel,ij} \equiv \frac{1}{2} \mu \left | \left (\vec{v}_i - \vec{v}_j \right ) \cdot \vec{r}_{ij} / r_{ij} \right |^2$. 
$\mu$ is the reduced mass of $i$ and $j$. 
For ModelIV $E_c = \epsilon c$, and $r_c = \gamma _2$ 
(see Table~\ref{tab:JR})\@. For ModelI $E_c$ is the peak of the inner spline 
on the $V_{vdW}$ term and $r_c$ 
is the location of that peak (see Fig.~\ref{fig:vdW_cmprsn}). 
It can be shown that for ModelI there exists no minimum above and within these 
cutoff 
values for all values of $\overline{B}_{ij}$. To be considered part of a 
dimer, a particle must be bonded to only one other particle that is bonded to 
no other. If the particle types are the same, it is a product dimer; if 
different, 
a reactant dimer. To be part of a cluster, a particle must be bonded to 
either multiple particles or to one other that is bonded to multiple 
particles. To be dissociated, a particle must be bonded to no other.  

\begin{figure}
\includegraphics*[width=\linewidth]{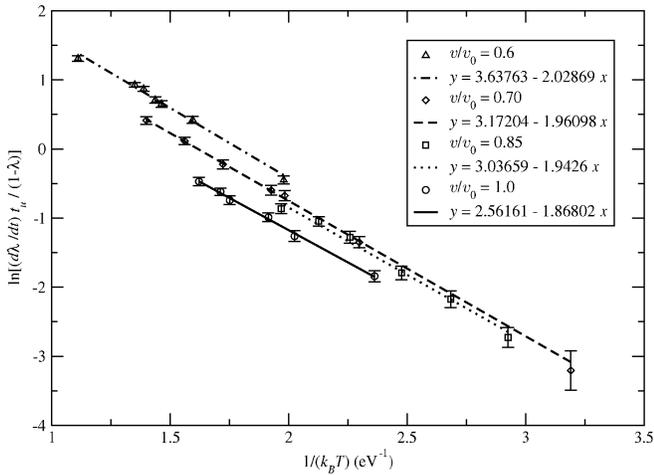}
\caption[Reaction rate of ModelIV for various internal energies and volumes 
vs.~temperature.]
{Reaction rate of ModelIV for various internal energies and volumes 
vs.~temperature. Lines are fits to the Arrhenius 
form for constant volume. Their slopes are the negative of the activation 
energy. $t_u$ is a unit of time and equals 10.180505 fs.}
\label{fig:ArrJR}
\end{figure}

\begin{figure}
\includegraphics*[width=\linewidth]{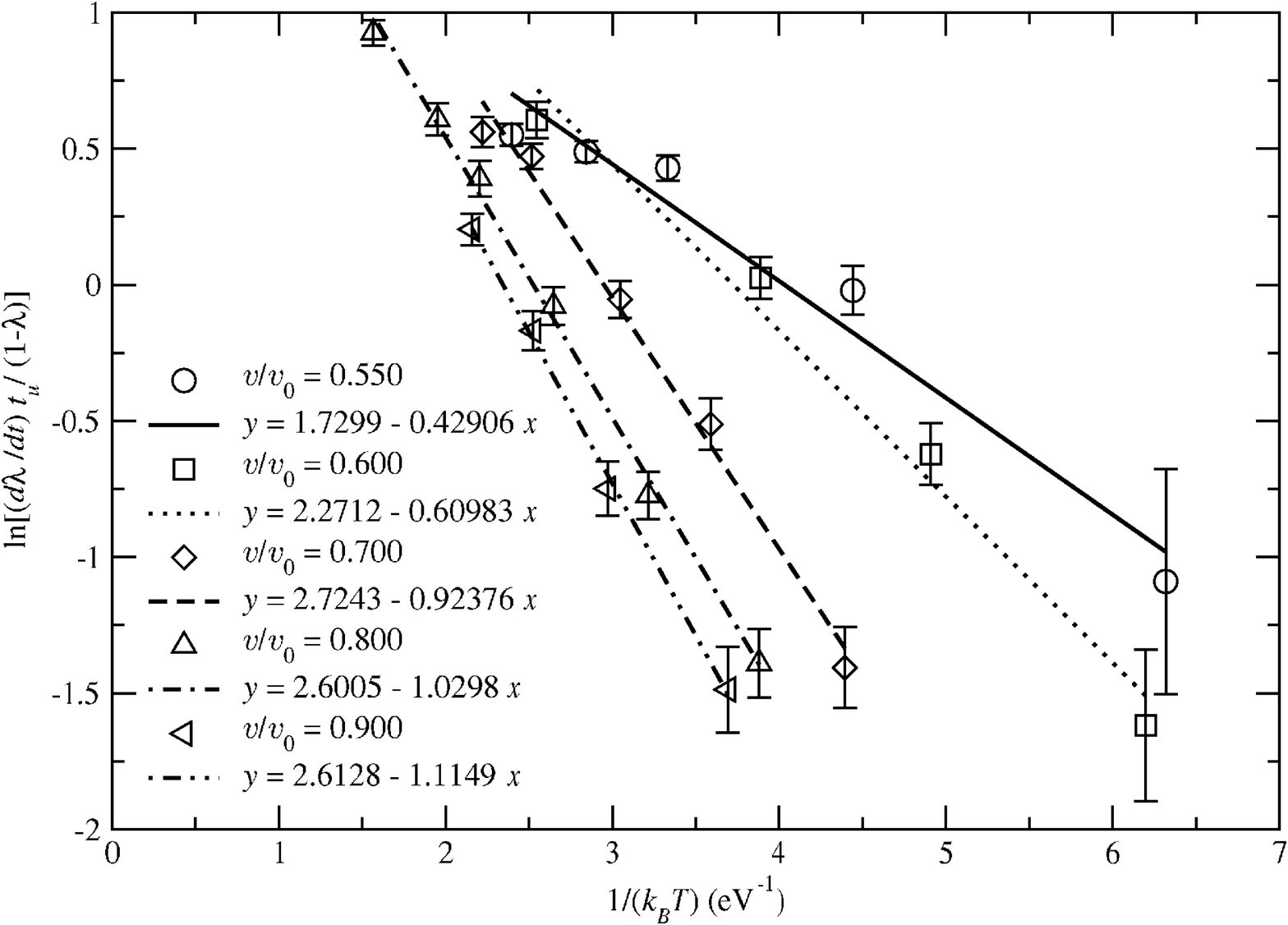}
\caption[Reaction rate of ModelI for various internal energies and volumes 
vs.~temperature.]
{Reaction rate of ModelI for various internal energies and volumes 
vs.~temperature. Lines are fits to the Arrhenius 
form for constant volume. Slope is the negative of the activation energy. 
$t_u$ is a unit of time.}
\label{fig:ArrBREW}
\end{figure}

With a bond defined, cookoffs that will reveal the reaction rate 
and the activation energy ($E_a$) for each model can be simulated. 
This is done for two reasons, to compare ModelIV to continuum models and to 
show another fortuitous difference between ModelI and ModelIV\@. 
The cookoffs are NVE 
simulations started at different values of compression and internal energy 
but all with the initial AB chemical composition. 
As a simulation is 
started, thermal energy is partitioned among the different modes. After this 
has occurred and the reaction has progressed somewhat, the reaction rate is 
sampled over a short time and the temporal average of the rate and its 
standard error are recorded for the current simulation. 
In Fig.~\ref{fig:ArrJR} the data are plotted for a series of simulations 
using the ModelIV potential. 
For each value of $v$, the data are fit to an Arrhenius reaction rate of the 
form~\footnote{
$(\lambda _e - \lambda)$ makes a more accurate coefficient, where $\lambda _e$ 
is the equilibrium value of $\lambda$ and could be a 
source of further temperature and density dependence.}
\begin{equation}
\label{eq:reac_rate} 
\dot{\lambda}=(1-\lambda )A\exp\left [\frac{-E_a} {k_B T}\right ], 
\end{equation}
where 
$\lambda$ is the degree of reaction defined computationally as the ratio 
of reacted (defined as above) material to total amount of material. 
$\dot{\lambda}$ is the reaction rate and $A$ is the frequency factor. 
The analogous plot for ModelI can be found in Fig.~\ref{fig:ArrBREW}. 

When contrasting the two figures, one 
notices that, over a similar range of volumes, ModelIV maintains a better fit 
to the Arrhenius form than ModelI and that the calculation of $E_a$ remains 
relatively constant when compared to the $E_a$ of ModelI\@. A probable 
contribution to the former observation is the manner in which a reaction is 
defined. At compressions typical in detonations, ModelI has a larger number of 
atoms in clusters and free states, which are not counted as reacted. 
Only stable dimers are. One may consider a cluster of two A atoms tightly 
bonded together with a third, B-type particle loosely bonded to one of them 
as a state more reacted than one in which A is tightly bonded to B and loosely 
bonded to the other A\@. However, by our definition of reaction, all particles 
in a cluster are considered unreacted and only contribute to the denominator of 
the calculation of $\lambda$.

$E_a$ for ModelI decreases with decreasing $v$. This is 
because at higher compressions an atom can have more neighbors within the 
bonding distance than with ModelIV\@. 
The bond-order coefficient then lowers the attraction 
between it and its neighbors, making it easier to break apart any existing 
bonds. For ModelIV, the repulsive core keeps neighbors away from dimers so that 
the bond-order coefficient and, thus, the dissociation energy remain 
unaffected by the same level of compression that would otherwise cause ModelI's 
dissociation energy to be reduced. For ModelIV there is a higher tendency 
than for 
ModelI for the compression to use up the space between dimers than that between 
the dimers' constituents. This is because for the former case a third particle 
would have to climb the repulsive core that was added to the $V_{vdW}$ term 
before it got within the range of the cutoff function $f_c$, where it can 
diminish the bond-order coefficient $\overline{B}_{ij}$. In the latter 
case a third particle would only need to overcome the barrier of ModelI's inner 
spline (see Fig.~\ref{fig:vdW_cmprsn}), which is small compared to the 
temperature for most situations 
considered in reaction and detonation. For ModelIV the activation energy goes 
up slightly with decreasing volume. The exact reason for this is not known at 
present, but simulations with ReaxFF on the high explosive RDX also show this 
trend~\cite{Strachan}\@. $E_a$ for ModelIV is roughly 1~eV greater than for 
ModelI\@. This is commensurate with the height of the repulsive core. 

\begin{figure}
\includegraphics*[width=\linewidth]{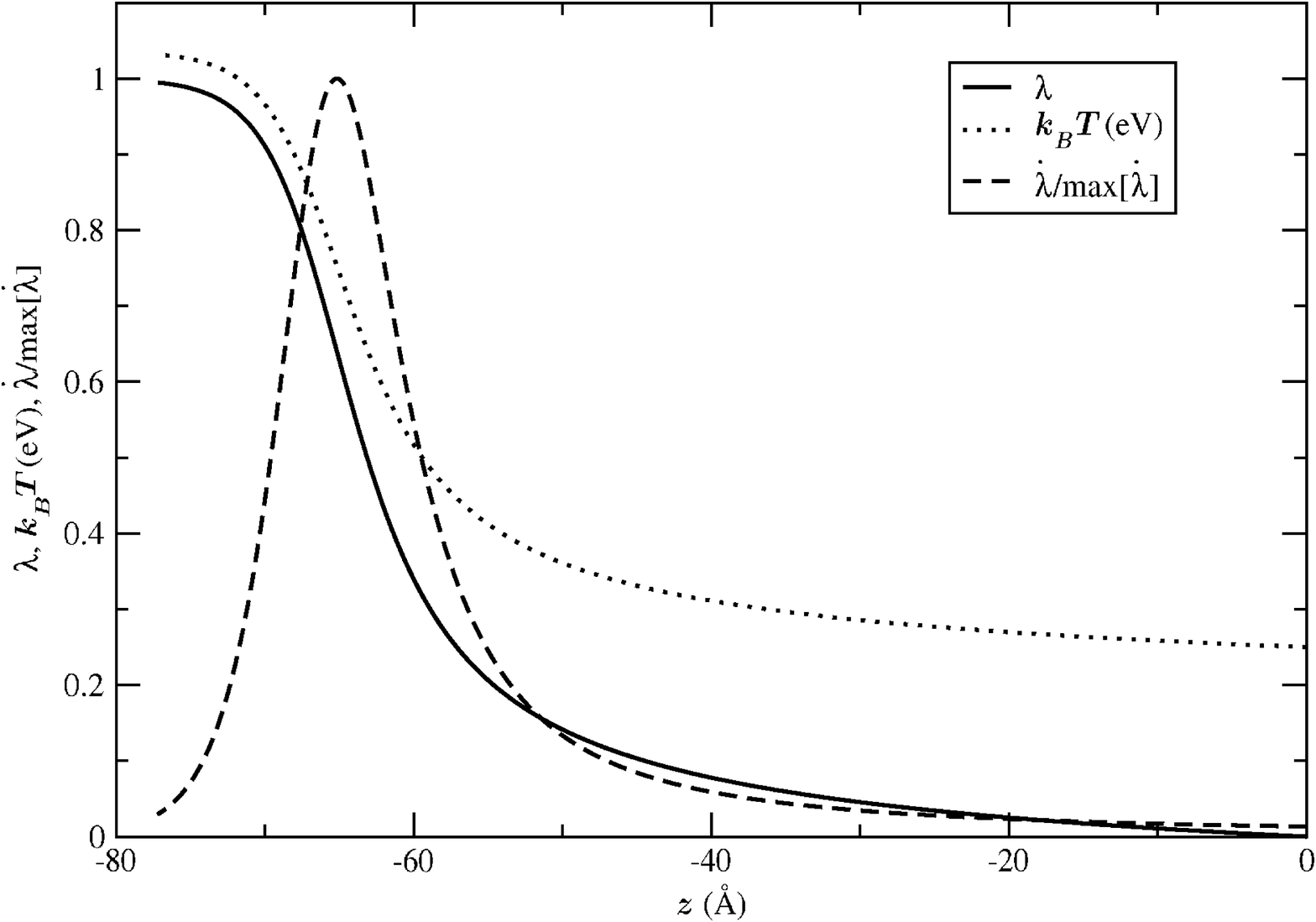}
\caption[Theoretical profiles for the degree of reaction ($\lambda$), 
temperature ($T$), and reaction rate ($\dot{\lambda}$).]
{Theoretical profiles for the degree of reaction ($\lambda$), temperature 
($T$), and reaction rate ($\dot{\lambda}$) for ModelIV, 
assuming a constant pressure and density. $\max[\dot{\lambda}]=8.94$~THz.}
\label{fig:lamprof}
\end{figure}

Even though we show that ModelIV fits an Arrhenius reaction rate in NVE cookoff 
simulations, we would like to know that this is the type of reaction that takes 
place in the reaction zone. 
We can use that reaction rate along with the 
von Neumann spike (vNs) state to predict a $\lambda$ profile. Since we do not 
have the full EOS 
throughout the reaction zone, we make some assumptions that will tend to 
shorten the width of the profile. Let's assume that $u_p$ increases linearly 
and that $v$ and $P$ are constant such that $T(\lambda )$. We conduct 
a series of 
shock tube simulations of a substance similar to ModelIV except that $Q$ is 
set to zero. We find the value of piston velocity that produces the constant 
zone the state of which falls on the theoretical $\mathcal{R}$ from 
Fig.~\ref{fig:Pv_REBO_JR}. This is the vNs state. The value of the specific 
volume at the vNs state is $v_n \approx 0.558~v_0$ and temperature 
$T_n \approx 0.25/k_B$. From Fig.\ref{fig:ArrJR}, 
we estimate the reaction rate  parameters (see Eq.~\ref{eq:reac_rate}) 
at the vNs state to be $A_n \approx 3.6$ and $E_{an} \approx 2.0$. 
Inserting the following equation for temperature 
\begin{equation}
\label{eq:Temp}
T(\lambda)=T_n + (T_j-T_n)\frac{\lambda}{\lambda_j}
\end{equation}
into Eq.~\ref{eq:reac_rate} and using the change of variable 
\begin{equation}
\label{eq:t2x}
\frac{d\lambda}{dz}=\frac{d\lambda}{dt}\frac{dt}{dz}=\dot{\lambda}\left(u_{sj}-\frac{u_{pj}}{2}\right)^{-1},
\end{equation} 
we generate a profile by iterating backward in space via a third order 
Runge--Kutta method (see Fig.\ref{fig:lamprof}). The variables with the 
subscript $j$ are the CJ values listed in Table~\ref{tab:CJ_JR}.  

The theoretical $\lambda$ profile reaches $\lambda_j$ at $z\approx -70$~\AA, 
much 
closer to the front than the $-700$~\AA~that we measure for the CJ value of the 
particle velocity in Fig.~\ref{fig:vz_ovrlp_crit_JR}. 
This does not prove that the 
reaction rate is Arrhenius in the reaction zone, but it does show that the 
order of 
magnitude of the measured reaction zone width is significant compared to 
the lower bound of the theoretical width. Had it been otherwise, this would 
have shown 
the reaction rate to not be Arrhenius and thus a different form, for example, 
a flame front. 

Replacing the assumptions in the analysis with more accurate ones, for 
example, variable $v$, would tend to lengthen the theoretical width because 
the expansion of the material would tend to cool it and slow the reaction 
rate. In lieu of the full EOS, we could have used data from an NEMD to fill 
in the state within the reaction zone. However, we would only be testing 
internal consistency. This comparison also suffers from an assumption of 
steady-state  
one-dimensionality. We have seen that the NEMD simulations are 2D\@. As we 
show in the companion paper, transverse waves traverse the reaction zone. 
This causes 
the profile to be time and $x$ dependent. The EOS throughout the reaction 
zone is currently being calculated for normal mode stability analysis 
\cite{Stewart}, 
which assumes an Arrhenius reaction rate. 

\section{Conclusion}

In this paper it was shown that the changes to the ModelI version of the REBO 
potential have 
reduced the amount of clustered and dissociated atoms at the CJ state (by RDF) 
if not throughout the reaction zone (by a snapshot of an NEMD simulation)\@. 
The CJ state of ModelIV is at a less compressed state and ModelIV's reaction 
zone is 
wider than ModelI's is. ModelI's unsupported detonation velocity is much better 
predicted by CJ theory, where ModelIV's propagates significantly faster. 
In the companion paper we investigate ModelIV's relative disparity more 
closely. 
The cookoff simulations reveal that ModelIV fits an Arrhenius 
reaction rate through a wider domain of compressions than does ModelI and it 
has a higher and less density-dependent activation energy. 

It was put forth that this higher 
$E_a$ causes the final state to be at a weak point since it might make more 
significant in the reaction rate the endothermic step of dissociation. Given, 
however, the 2D shape of the front in Fig.~\ref{fig:snpsht_JR_frnt}, such a 
1D explanation is probably insufficient. Higher $E_a$s should increase the 
likelihood of 1 and 2D instabilities in detonation waves \cite{Lee,Stewart}. 
The current value of $\approx 2$~eV is consistent with the presumed activation 
energies for many conventional HEs. 

The goal of the current 
research is the improvement of REBO, which we believe we have achieved, 
mainly, by introducing increased atomic repulsion, which reduces 
the reaction cross section. It was shown that ModelIV's reaction zone is wider 
than 
with ModelI and that the CJ state is a molecular one. REBO's insensitivity to 
initiation is increased, and a thicker induction zone is introduced. The 
reaction rate now behaves in a more Arrhenius manner, and the 
product EOS behaves more like a polytropic gas as indicated by the hyperbolic 
shape of its equilibrium $\mathcal{H}$. Some measurements, 
for example, the adiabatic $\gamma$---as is shown in the companion paper---and 
the amount of compression at CJ, are 
more commensurate with conventional explosives~\cite{Mader}. However, some are 
not too close, for example, the reaction temperature and shock velocity. 3D 
simulations are needed to investigate how the changes truly compare to real 
experiments.  

\acknowledgments
The authors would like to thank Sam Shaw, Alejandro Strachan, Brad Holian, 
Tommy Sewell, Yogesh Joglekar, 
and David Hall for useful conversations. This material was 
prepared by the University of California under Contract W-7405-ENG-36 and 
Los Alamos National Security under Contract DE-AC52-06NA25396 with the 
U.S. Department of Energy. The authors particularly wish to recognize funding 
provided through the ASC Physics and Engineering Modeling program.

\end{document}